# Reference-free single-point holographic imaging and realization of an optical bidirectional transducer

Seungwoo Shin[1,2], KyeoReh Lee[1,2], YoonSeok Baek[1,2], and YongKeun Park[1,2,3,*]

[1]*Department of Physics, Korea Advanced Institute of Science and Technology, Daejeon 34141, Republic of Korea.*
[2]*KAIST Institute for Health Science and Technology, KAIST, Daejeon 34141, Republic of Korea.*
[3]*Tomocube, Daejeon 34051, Republic of Korea.*

One of the fundamental limitations in photonics is the lack of a bidirectional transducer that can convert optical information into electronic signals or *vice versa*. In acoustics or at microwave frequencies, wave signals can be simultaneously measured and modulated by a single transducer. In optics, however, optical fields are generally measured via reference-based interferometry or holography using silicone-based image sensors, whereas they are modulated using spatial light modulators (SLMs). Here, we propose a scheme for an optical bidirectional transducer using an SLM. By exploiting the principle of time-reversal symmetry of light scattering, two-dimensional reference-free measurement and modulation of optical fields are realized. We experimentally demonstrated the optical bidirectional transducer for optical information consisting of 128×128 spatial modes at visible and short-wave infrared wavelengths.

## I. INTRODUCTION

The ability to measure or modulate both the amplitude and phase information of a light field is central to optical metrology, with potential applications in materials science, nanotechnology, and biophotonics [1-5]. Optical phase information can be obtained indirectly by recording the patterns that are formed as a result of the interference of a sample beam with a well-defined reference beam [Fig. 1(a)]. Interference-based holographic imaging techniques [6-8] have led to the emergence of various research disciplines, and their applications have been further expanded with recent advances in the development of silicon image sensors, such as charge-coupled devices and complementary metal-oxide-semiconductor devices.

However, conventional holographic imaging techniques require both of an interferometer and an image sensor, and this requirement greatly restricts realization of an optical bidirectional transducer and broader application of holographic imaging, particularly at wavelengths where the applicability of high-quality image sensors is limited. Radio or acoustic waves can be measured and modulated using bidirectional transducers such as an antenna [Fig. 1(b)] [9,10]. On the contrary, the measurement and modulation of optical fields are achieved using separate principles: silicon-based image sensors are used to measure the interference patterns of the sample and reference beams that are generated in interferometry [1-3], whereas spatial light modulators (SLMs) are used to modulate optical fields through the reorientation of liquid crystal molecules or the actuation of deformable surfaces or micro reflective elements [5,11]. The simultaneous measurement and modulation of optical fields using a single electronic device or an optical bidirectional transducer have not yet been demonstrated.

Here, we propose and experimentally demonstrate a scheme for an optical bidirectional transducer using an SLM and a single-point detector [Fig. 1(c)]. Instead of employing an interferometer and an image sensor, the proposed optical bidirectional transducer measures and modulates wide-field optical fields by exploiting time-reversal symmetry of light scattering and the original functionality of an SLM, respectively.

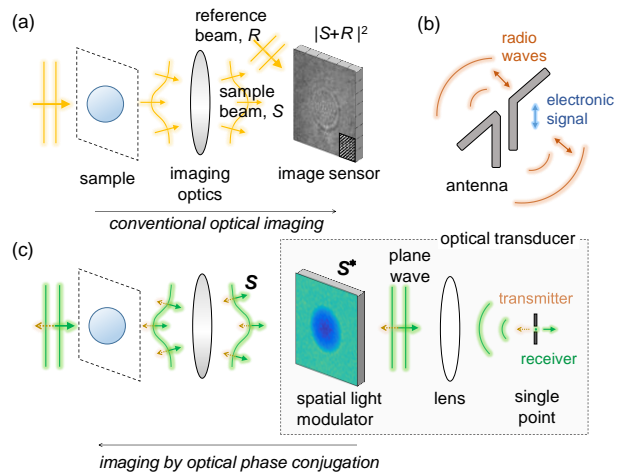

FIG. 1. Lack of an optical bidirectional transducer. (a) Schematic illustration of conventional digital holography: an image sensor records an interference pattern between a sample beam and a reference beam. (b) Radio waves can be simultaneously measured and modulated by a single transducer. (c) Schematic illustration of the proposed optical bidirectional transducer: all optical power in the incident field $S$ converges to a single point when the field is modulated by its optical phase conjugation $S^*$. Using the original functionality of a spatial light modulator, optical fields can also be modulated by the optical bidirectional transducer.

## II. THEORY OF OPERATION

The principle of measuring optical fields is based on time-reversal symmetry of light scattering [12,13]. After a light-matter interaction, plane-wave illumination results in a scattered field $S$ [Fig. 1(c)]. If the scattered field is propagated back to the sample in a time-reversed manner, the beam will become a plane wave after a light-matter interaction [12,13]. To use this time-reversal symmetry for measuring an incident field, we exploited time-reversal nature of optical phase conjugation. The optical phase conjugation of a monochromatic wave in the spatial domain is identical to the time reversal of the wave, $E^*(r,t) = E(r,-t)$ [12,14]. Thereby, a scattered field $S$ can be rewound to a plane wave by modulating the scattered field using its optical phase conjugation $S^*$ [Fig. 1(c)]. By sequentially displaying complex-valued patterns on an SLM, the intensities of the modulated fields are measured by a point detector. Then, from the time-reversal discussion, the displayed pattern corresponding to the maximum intensity can be uniquely determined as the optical phase conjugation of the incident field, from which both amplitude and phase images of the field can be produced.

For measuring optical fields in practice, it is crucial to effectively find the pattern which gives the maximum focused intensity after a lens. Recently, several algorithms have been developed to achieve point optimization [15-17]. In this work, we utilized basis transformation. By recording and analyzing responses of incident field modulated by a set of complex-valued patterns, we can find the pattern maximizing focused intensity at a point. However, the present approach is not limited to this basis transformation method; other methods including compressive sampling approaches can also be utilized for finding the pattern maximizing focused intensity at a point.

When an incident field $S$ is modulated by a complex-valued map $D_n$, the intensity measured by a single-point detector located at the focus of a lens can be expressed as $I_n = |(D_n \circ S)_{|k|=0}|^2 = |\int D_n \circ S \, da|^2$, where $\circ$ represents element-wise multiplication or the Hadamard product, $k$ is a spatial frequency vector, and $\int da$ denotes a surface integral. For effective working, we construct a set of displaying patterns using known phase shifts $e^{i\phi_p}$ and an arbitrary basis $H$ as follows: $D_n = e^{i\phi_p} H_q + H_1$, where $i = \sqrt{-1}$, $p \in \{1,2,3\}$, and $q = 1$–N, where N represents number of basis vectors in the basis $H$ or number of spatial modes. Then, the intensities measured by a single-point detector are equal to

$$I_n = \left| e^{i\phi_p} \int H_q \circ S \, da + \int H_1 \circ S \, da \right|^2 \\ \equiv \left| e^{i\phi_p} s_q + r \right|^2 \quad (1)$$

where $s_q \equiv \int H_q \circ S \, da$ represents the decomposition coefficient of $S$ calculated on a basis vector $H_q$ and $r \equiv \int H_1 \circ S \, da$ is a constant. By regarding the constant value $r^*$ as a global phase and using the known phase shifts, the complex values $s_q$ for all $q$ can be obtained from Eq. (1). Then, the incident field in the standard basis can be expressed via a basis transformation: $S(x,y) = \sum s_q H_q(x,y)$.

## III. EXPERIMENTAL RESULTS

For experimental demonstrations, we used three optical components: a digital micromirror device (DMD) as an SLM, a lens, and a single-mode fiber (Fig. 2). A DMD consists of up to a few million micromirrors, which are individually switchable between the on and off states at a speed of tens of kHz. We utilized a DMD rather than a liquid-crystal SLM to achieve fast modulation of the light field and broadband operation. However, the proposed method is not limited only to this specific type of light modulator; any type of SLM, regardless of its amplitude or phase modulation, can be used.

### A. Optical bidirectional transducer: demonstration in visible wavelength

The workflow of single-point holographic imaging or an optical bidirectional transducer as a receiver is presented in Fig. 2. An incident field is sequentially modulated by multiple binary patterns displayed on a DMD [Figs. 2(a) and 2(b)]. The intensities of the modulated fields are then measured by a single-point detector [Fig. 2(c)]. To construct the set of displaying patterns, Hadamard basis was used [Fig. 2(d)] [18]. In addition, for displaying complex values on the DMD, we employed the superpixel method [19] (details in Appendix A). From the measured intensities by the single-point detector, the incident field can be retrieved using Eq. (1) [Fig. 2(e)].

We first experimentally validated the proposed optical bidirectional transducer for measuring optical fields (Fig. 3). An interconvertible setup was used for the direct comparison of the proposed optical bidirectional transducer and conventional digital holography using an off-axis Mach-Zehnder interferometer (details in Appendix B). To demonstrate the feasibility of our method for various types of samples, we measured the fields diffracted by a phase object [a polystyrene (PS) bead with a diameter of 10 μm, Figs. 3(a) and 3(b)] and an amplitude object [the number 7 representing group 7 from the United States Air Force (USAF) resolution test chart, Figs. 3(c) and 3(d)]. Both the amplitude and phase images of the samples that were measured using the proposed method were well consistent with those obtained via conventional holographic imaging, thereby serving as a proof of principle for the optical bidirectional transducer as a receiver.

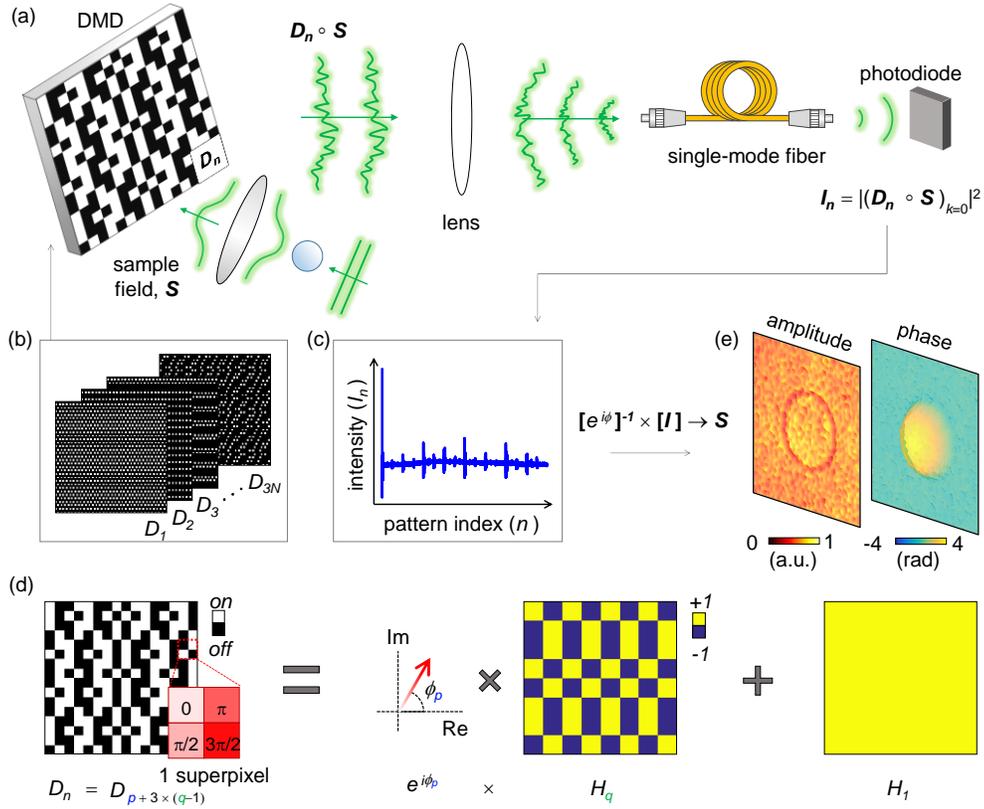

FIG. 2. Workflow of the proposed optical bidirectional transducer when measuring optical fields. (a) The optical field $S$ that is diffracted by a sample is modulated by a binary pattern $D_n$ displayed on a digital micromirror device (DMD). The intensity of the modulated field at a single point is measured by a photodiode after passing through a lens and a single-mode fiber. (b) Multiple binary patterns that are sequentially displayed on the DMD. (c) Measured intensities corresponding to the displayed pattern index. (d) Construction of binary patterns using phase shifts and a Hadamard basis. $e^{i\phi_p}$ and $H_q$ denote the phase shift ($p \in \{1, 2, 3\}$) and the $q$th basis vector of the Hadamard basis ($q = 1$–$N$), respectively. The superpixel method of displaying a complex-valued map using a DMD is illustrated in the inset. (e) The measured incident field. The narrow arrows indicate the entire sequence of an optical bidirectional transducer as a receiver for measuring incident optical field information.

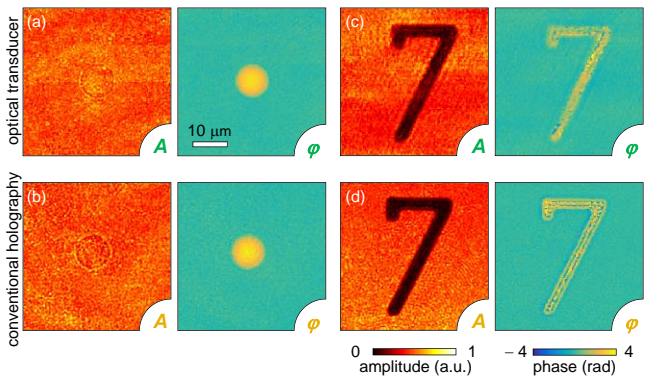

FIG. 3. (a to b) The field diffracted by a phase object, a 10-μm-diameter polystyrene bead immersed in oil, is measured by the optical bidirectional transducer (a) and conventional holography (b). (c to d) The field diffracted by an amplitude object, the numeral "7" representing group 7 in the United States Air Force resolution test chart, is measured by the optical bidirectional transducer (c) and conventional digital holography (d). The amplitude and phase images are labelled with the symbols $A$ and $\varphi$, respectively, in the bottom right corner of each figure.

In order to experimentally demonstrate the proposed optical bidirectional transducer, we designed an experiment as presented in Fig. 4. First, the optical bidirectional transducer measures a diffracted field from a cluster of 10–μm–diameter PS beads immersed in oil [Figs. 4(a) and 4(b)]. Next, to verify the capability of accurate measurement and modulation of optical fields using the proposed approach, the phase conjugation of the measured field is illuminated to the same cluster of the beads [Fig. 4(c)]. By the time-reversal symmetry of light scattering, the transmitted wave becomes a plane wave after diffraction from the cluster. Then, the plane wave is measured and verified using an interferometer, which clearly shows the successful working of the optical bidirectional transducer [Fig. 4(d)]. In addition, for purposes of comparison, a diffracted field by the cluster of the beads under plane-wave illumination is measured by the conventional interferometer [Figs. 4(e) and 4(f)]. In this work, to modulate optical fields using a DMD, amplitude holograms are displayed using time-multiplexing method [20].

## B. Single-point holographic imaging at short-wave infrared wavelength

The proposed scheme for an optical bidirectional transducer offers several advantages. Because it does not require interferometry and an image sensor for measuring wide-field optical fields, the proposed method allows to robust measurement of optical fields, without undesired random fluctuations in interferometry and image deteriorations caused by read-out noise (details in Appendix C) [21]. Most importantly, the principle presented in this work can be readily applied to electromagnetic waves of other wavelengths, ranging from X-ray and deep ultraviolet wavelengths to infrared and terahertz wavelengths.

To further demonstrate the applicability of the proposed method at other wavelengths, we used the optical bidirectional transducer as a receiver in the short-wave infrared (SWIR) at a wavelength of 1.55 μm (Fig. 5). At SWIR wavelengths, silicon image sensors are blind; as an alternative, indium gallium arsenide (InGaAs) image sensors can be utilized, but their applications are highly limited because of their limited response and pixel resolution as well as their high price. A silicon wafer etched with patterns was used as the phase object for the SWIR experiment [Fig. 5(a)]. A laser with a wavelength of 1.55 μm and an InGaAs photodiode were employed [Fig. 5(b)], and the field diffracted from the wafer was measured using the optical bidirectional transducer [Figs. 5(c) and (d)]. The height map $d$ of the wafer was calculated from the measured phase image $\phi(x,y)$ [Fig. 5(d)] as $d(x,y) = \lambda \cdot \Delta\phi(x,y) / (2\pi \Delta n)$ where $\Delta n$ represents the difference between the refractive indices of silicon and air at the wavelength of 1.55 μm.

To verify the SWIR field measurement using the proposed method, a topographic map of the letter "P" was measured via atomic force microscopy (AFM). For direct comparison, a magnified image of the height map obtained using the proposed method is shown alongside the topographic map measured via AFM [Fig. 5(e)]. To systematically compare these images, the measured profiles are presented on the same plot [Fig. 5(f)]; the results serve as validation of the field image measured in the SWIR.

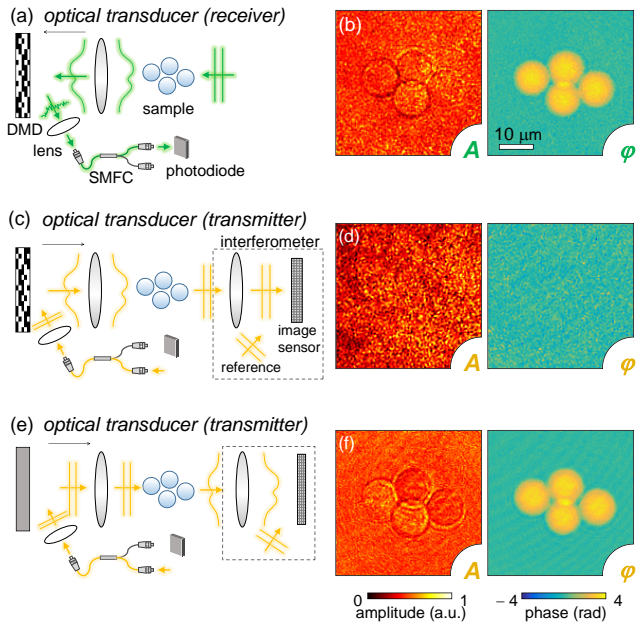

FIG. 4. Experimental demonstrations of an optical bidirectional transducer. (a) The optical bidirectional transducer measures an optical field diffracted by a cluster of 10-μm-diameter polystyrene beads immersed in oil. SMFC; single-mode fiber coupler. (b) The amplitude and phase images of the measured field. (c) The optical bidirectional transducer modulates an optical field which is the phase conjugation of the measured field. After a light matter interaction, the modulated wave becomes a plane wave which can be verified using a conventional interferometer. (d) The amplitude and phase images of the plane wave measured by the interferometer. (e) For comparison, the optical bidirectional transducer transmits a plane wave, and a diffracted field is measured by the interferometer. (f) The amplitude and phase images of the diffracted field.

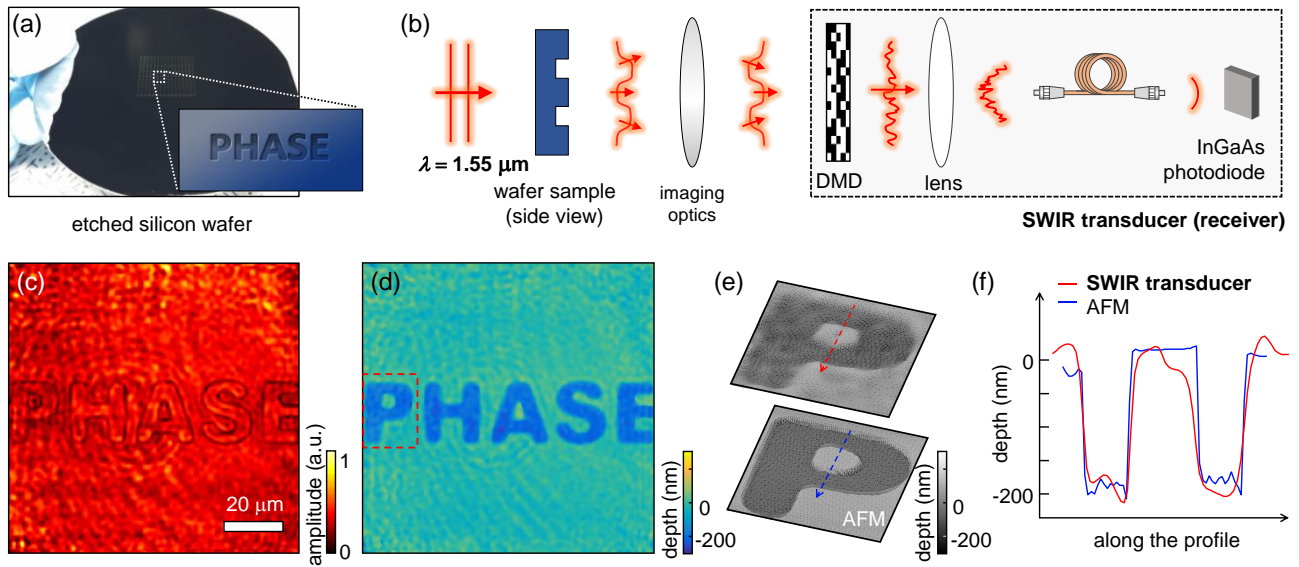

FIG. 5. Single-point holographic imaging in the short-wave infrared, where conventional silicon detectors are blind. (a) A photograph of a silicon wafer, which acts as a phase object at a wavelength of 1.55 μm. On the wafer, repetitive patterns are etched to a depth of 200 nm in the form of the word "PHASE." (b) A beam whose wavelength is 1.55 μm transmitted through the wafer is projected onto a DMD. The field diffracted by the pattern is measured by the optical bidirectional transducer. For visualization purposes, the DMD, which is in fact a reflective modulator, is depicted as a transmissive modulator. (c) The amplitude image of the measured field. (d) The depth map of the wafer produced from the phase image of the measured field. (e) A magnified view of the portion of the depth map indicated by the dotted red box in (d) (upper) and a topographic map of the letter "P" (lower) as measured via atomic force microscopy. (f) The depths along the profiles indicated in (e).

It should be emphasized that the present approach is fundamentally different from previously reported methods using single-pixel cameras in interferometry [22-29]. Conventional single-pixel cameras can only measure intensity images from intensity-correlation measurements. For retrieving phase information of light, the single pixel cameras employ interferometry to record interference patterns or diffracted patterns. By analyzing the recorded interference fringes or diffraction patterns, the lost phase information in photographs is retrieved. These previous works using single-pixel cameras in interferometry are essentially identical to using image sensors in interferometry.

Contrary to the typical single-pixel cameras in interferometry, the key innovation and novelty of this work are showing the unique determination of the pattern maximizing the focused intensity at a point as an optical phase conjugation using the time-reversal symmetry of light scattering, and utilizing this idea to holographic imaging. By exploiting the super-pixel method [19] to a DMD, incident field is directly modulated by complex-valued patterns for finding the pattern maximizing the focused intensity at a point, not for recording interference patterns or diffracted patterns (additional proof in Appendix D).

Using the present approach for holographic imaging, we can experimentally realize an optical bidirectional transducer. Similar to applications of ultrasonic waves in medical diagnosis and treatment, we expect that the proposed concept for an optical bidirectional transducer can be applied to various areas, such as optical manipulation and measurement of biological cells which are now performed by employing a spatial light modulator and interferometric imaging, respectively [30]. Also, unlike existing methods of wavefront measurement and shaping [5,16,31,32] and digital optical phase conjugation [33-36], our method allows to both of measure and modulate wide-field optical fields using a single device, and does not require an interferometer and an image sensor, thereby its applicability is significantly broader.

## IV. CONCLUSION AND PROSPECTS

In this work, we have proposed and experimentally demonstrated scheme for an optical bidirectional transducer that can convert optical information into electronic images or vice versa. By exploiting time-reversal symmetry of light scattering, we have shown unique determination of the pattern maximizing focused intensity at a point as an optical phase conjugation, and it is used for a new approach for holographic imaging without the use of an interferometer and an image sensor. Furthermore, the applicability of the proposed method in the SWIR was verified by measuring complex field images at the wavelength of 1.55 μm.

From a technical perspective, the proposed method can be combined with illumination engineering methods to realize holographic imaging with sub-diffraction resolution using the synthetic aperture method [37,38] or 3-D refractive index (RI) tomography exploiting optical diffraction tomography [39-41]. Moreover, the optical bidirectional transducer as a receiver can be extended to spectroscopic holography using the linear dispersion of a DMD and a spectrometer. In this work, 49,152 patterns were displayed on the DMD with a display rate of 10 kHz, from which we acquired an optical field in 4.9 seconds. Although the present method has limitations with regard to dynamic studies because of the need for sequential measurements, algorithms for compressive sensing [21,23] can be adapted to improve the image acquisition rate. We expect that the proposed method may offer solutions for wavefront shaping, adaptive optics, and high-fidelity holographic imaging at wavelengths where the applicability of high-quality image sensors is limited.

## ACKNOWLEDGMENTS

We thank Hwan-Seop Yeo (KAIST) for assisting the use of AFM. This work was supported by KAIST, BK21+ program, Tomocube, and National Research Foundation of Korea (2015R1A3A2066550, 2014M3C1A3052567, 2014K1A3A1A09063027).

## APPENDIX A: SUPERPIXEL METHOD AND PATTERN CONSTRUCTION

To display a complex-valued map on the DMD, we utilized the superpixel method proposed by the Mosk group [19]. With the appropriate placement of the single lens, linearly varying phase shifts can be assigned to each micromirror of the DMD, increasing by π and π/2 along the horizontal and vertical directions, respectively [inset of Fig. 2(d)]. Thus, the phase shifts of the 4 neighbouring micromirrors in a rectangular region will be equally divided between 0 and 2π, allowing superpixels to be defined as 2×2 arrays of DMD micromirrors. To allow the phase shifts in a superpixel to be combined, the setup should be designed such that the micromirrors in each superpixel are unable to be resolved by optics. Thus, by turning on different combinations of the micromirrors in a superpixel, 9 different complex values can be displayed.

The addition of $e^{i\phi_p} \times H_q$, the phase-shifted $q$-th basis vector of the N-dimensional Hadamard basis, to $H_1$, the first basis vector, generates a complex-valued map, $e^{i\phi_p} \times H_q + H_1$, where $p \in \{1,2,3\}$ and $q = 1$–N. Then, a binary pattern $D_n$ is generated from the complex-valued map via the superpixel method. Since a superpixel must be able to modulate 9 different complex values, the phase shifts are set to 0, π/2, and π for $p = 1, 2,$ and 3, respectively.

For the construction of the binary patterns, we used the Hadamard basis method with N = 128. With three phase shifts, 49152 patterns were displayed on the DMD with a display rate of 10 kHz, from which we acquired an optical field in 4.9 seconds. The image acquisition rate depends on the number of spatial modes N and the display rate $f$ of the DMD as $f / 3N^2$. Superpixels consisting of 2×2 arrays of DMD micromirrors were chosen; altogether, 256×256 micromirrors were used to display binary patterns in the visible wavelength range.

The numerical aperture of OL1 and the focal length of TL1 were appropriately selected to ensure that all micromirrors in each superpixel were unresolvable. In the SWIR range, because of the larger diffraction limit, each superpixel was chosen to consist of 4×4 DMD micromirrors, and in total, an array of 512×512 micromirrors was used to display the binary patterns.

## APPENDIX B: EXPERIMENTAL SETUP

To experimentally validate the optical transducer for measuring optical fields, we constructed an interconvertible set-up for the direct comparison of two imaging methods: optical transducer as a receiver and conventional digital holography using a Mach-Zehnder interferometer (Fig. 6). For a systematic comparison, a flip mirror and a 2×2 single-mode fiber optic coupler (2×2 SMFC; FC532-90B-FC, Thorlabs Inc.) were used to share the optical setup between the two methods. To allow both conjugate planes of the sample plane to be used as the image planes for the two different imaging methods, the optical setup was constructed symmetrically. The same visible laser ($\lambda$ = 532 nm; LSS-0532, Laserglow Inc.) was used as the source for both imaging methods.

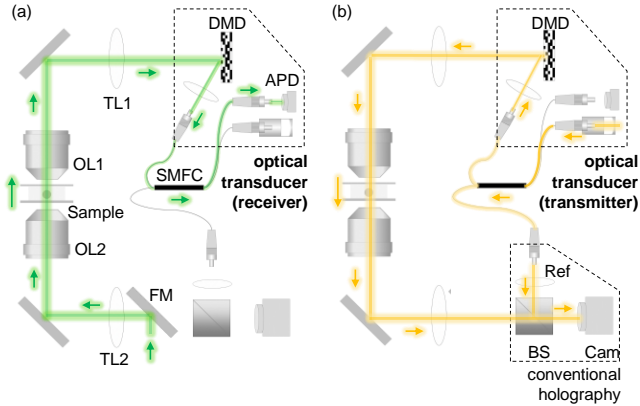

FIG. 6. Experimental setup for validation of the optical transducer. An interconvertible set-up for measuring and verifying optical fields diffracted by a sample using the optical transducer and conventional digital holography. (a) Measurement of the optical field using sinlge-point holography or the optical transducer as a receiver. (b) Measurement of the optical field via Mach-Zehnder interferometry for verifying modulated optical fields by the optical transducer. FM, flip mirror; 2×2 SMFC, 2×2 single-mode fiber optic coupler; OL, objective lens; TL, tube lens; APD, avalanche photodiode; BS, beam splitter; Cam, camera; Ref, a reference beam.

For the measurement of an optical field using the optical transducer, a plane wave was introduced into the optical setup by the flip mirror [Fig. 6(a)]. The plane wave impinged on the sample, and the beam diffracted by the sample was projected onto the DMD (maximum switching rate = 22.7 kHz; V-7001, Vialux Inc.) by an objective lens (OL1, 0.7 NA; LUCPLFLN60X, Olympus) and a tube lens (TL1, $f$ = 250 mm). Then, the optical field was sequentially modulated by multiple binary patterns, and an avalanche photodiode then measured the intensity of the light passing through a single lens and one arm of the 2×2 SMFC. Finally, the optical field of the diffracted beam could be retrieved from the measured intensities using Eq (1).

For comparison, the optical field diffracted by the same sample was measured via Mach-Zehnder interferometry [Fig. 6(b)]. The laser beam was split into two arms (sample and reference arms) by the 2×2 SMFC. To illuminate the sample with a plane wave, the optical transducer transmitted a planar wavefront. An objective lens (OL2, 0.7 NA; LUCPLFLN60X, Olympus) and a tube lens (TL2, $f$ = 180 mm) were used to project the beam diffracted by the sample onto a camera (FL3-U3-13Y3M-C, FLIR Inc.), where the sample beam interfered with a reference beam at a slightly tilted angle to generate an off-axis hologram. From the measured hologram, the optical field diffracted by the sample was retrieved via the Fourier transform method [42].

To demonstrate the broadband capability of our method, a SWIR laser beam ($\lambda$ = 1.55 μm; SFL1550P, Thorlabs Inc.) was directed onto an etched wafer sample by a 4-$f$ telescopic system comprising a tube lens (TL2) and an objective lens (OL2, 0.3 NA; LCPNL10XIR, Olympus). The beam diffracted from the wafer was collected and projected onto the DMD by another 4-$f$ telescopic system comprising a tube lens (TL1) and an objective lens (OL1, 0.65 NA; LCPNL50XIR, Olympus). The intensity of the light passing through a single-mode fiber (P3-SMF28E-FC-2, Thorlabs Inc.) was measured by an InGaAs switchable gain-amplified detector (PDA10CS-EC, Thorlabs Inc.).

For measuring various samples, we used PS beads, USAF resolution test chart, and an etched double-sided polished wafer. As a phase object at visible wavelength, PS beads with a diameter of 10 μm ($n$ = 1.5983 at $\lambda$ = 532 nm, Sigma-Aldrich Inc.) immersed in index-matching oil ($n$ = 1.5660 at $\lambda$ = 532 nm, Cargille Laboratories). To separate aggregated beads, the beads in the immersion oil were sandwiched between coverslips before measurement. To measure the complex field diffracted by a phase object in the SWIR, we prepared a double-sided polished wafer on which repetitive patterns were dry etched.

## APPENDIX C: THEORETICAL COMPARISON TO INTERFEROMETRIC IMAGING METHODS

The principle of the proposed optical transducer as a receiver is different from other methods using a single-pixel camera in interferometry, in both conceptual and theoretical perspectives. From a conceptual viewpoint, the conventional interferometric imaging methods using a single-pixel camera records an interference image between a sample beam and a reference beam through sequential measurements, whereas the present method seeks a pattern which makes the maximum intensity at a point.

Theoretically, the present method exhibits superior performance compared to the interferometric methods using a single-pixel camera, because the present method has the inherent common-path geometry. Due to the common-path geometry, single-point holography avoids randomly fluctuating phases in an interferometer, which are inevitable in interferometric systems.

To be clear, we concretely describe the advantage. In an interferometer, undesired fluctuating phase $\varphi(t)$ between the sample and reference beams are inevitable due to random fluctuations in optics. Because of this random fluctuations, resulting interference signals between sample $S$ and reference $R$ beams can be written as

$$\left|R + S e^{i\varphi(t)}\right|^2 = |R|^2 + |S|^2 + R^* S \times e^{i\varphi(t)} + R S^* \times e^{-i\varphi(t)}, \quad (A1)$$

which represents temporally fluctuating bright and dark fringes in interference images. Thus, due to the randomly fluctuating phase, time-consuming records of an interference image could be corrupted, which deteriorates the fidelity to the field retrieval.

On the contrary, using the present method, the measured intensity and retrieved signal are $I_n = \left|e^{i\phi_p} s_q + r\right|^2$ and $\{s_q r^*\}$, respectively. Because the random fluctuating phase $\varphi(t)$ is included in both of $s_q$ and $r$, thereby, the random fluctuating phase becomes naturally canceled out in the retrieved terms $\{s_q r^*\}$. Due to the intrinsic elimination of the fluctuating phase, the present work theoretically shows superior performance for measuring optical fields.

# APPENDIX D: PROOF OF THE UNIQUE DETERMINATION USING CAUCHY-SCHWARZ INEQUALITY

For measuring optical fields using the proposed method, we have shown that a pattern, which maximizes the focused intensity after a lens, is identical to the optical phase conjugation of an unknown incident field, by exploiting the time-reversal symmetry of light scattering.

This unique determination can be proved again using another way, a fundamental theorem in linear algebra, Cauchy-Schwarz inequality for complex numbers. If $\{u_1, u_2, …, u_n\}$ and $\{v_1, v_2, …, v_n\}$ are complex numbers, then the inequality states that

$$\left|\sum_{i=1}^{n} u_i \times v_i^*\right|^2 \leq \sum_{j=1}^{n}|u_j|^2 \sum_{k=1}^{n}|v_k|^2, \quad (A2)$$

where $*$ denotes complex conjugation, and the equality holds when $\{u_1, u_2, …, u_n\}$ and $\{v_1, v_2, …, v_n\}$ are linearly dependent.

Figure 7 describes modulation of an unknown incident field $S$ by displaying a complex-valued pattern $D$ on a spatial light modulator (SLM). Because the complex-valued pattern is discretely displayed by pixels on an SLM, the intensity of focus at a point can be written as $I = \left|(S \circ D)_{k=0}\right|^2 = \left|\sum_{i=1}^{N} s_i \times d_i\right|^2$ where $i$, $s_i$, and $d_i$ denote the index of an SLM pixel and partial regions of the unknown field and the displayed pattern with respect to the indexed pixel, respectively. By comparing the above equations, the focused intensity at a point is maximized when the displayed complex-valued pattern linearly depends on the complex conjugation of the unknown field, i.e., $D_{max} = c S^*$ where $c$ is a constant complex number. Thus, the pattern, which maximizes the focused intensity, has the same amplitude distribution and the complex conjugation of the phase distribution of an unknown field.

Intriguingly, the pattern, which maximizes the intensity of focus, includes not only phase information but also amplitude information of an incident field. Even if an incident field has dark parts in the amplitude distribution, the maximizing pattern has the same amplitude distribution, not a constant amplitude distribution. Thereby, finding the maximizing pattern uniquely determines both of amplitude and phase distributions of an unknown field.

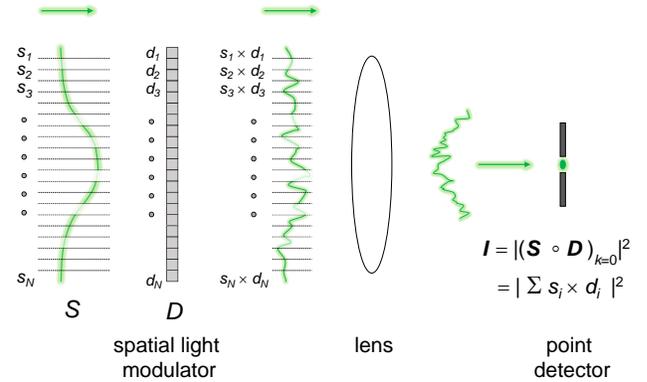

FIG. 7. Schematic of the optical transducer as a receiver with considering discrete modulation. An unknown field is discretely modulated on pixels of a spatial light modulator. The discretely modulated field is focused by a lens and the intensity of a focus at a point is measured by a point detector. The intensity can be expressed by the square of a sum of discrete modulations. We depicted nonuniform brightness of the visualized wavefront for considering an arbitrary field which has dark parts in the amplitude distribution.